\documentstyle[preprint,aps]{revtex}
\begin{document}
\draft
\title{The Gauge Technique\thanks{%
in memory of Abdus Salam, to whom we owe the gauge technique.} for Heavy
Quarks}
\author{R.~Delbourgo\cite{Author1} and D.~Liu\cite{Author2}}
\address{University of Tasmania, GPO Box 252C, Hobart,\\
Australia 7001}
\date{\today }
\maketitle

\begin{abstract}
It is possible to determine an off-shell propagator for heavy quarks to
order $1/m$ in mass and in any covariant gauge $\xi $, which applies
universally to all the quarks, by using the gauge technique. The result for
the leading behaviour of the propagator is 
\[
S(v\cdot k)=\Gamma (1+2\alpha _\xi )\frac{1+\gamma \cdot v}{2v\cdot k}\left(-%
\frac{v\cdot k}\Lambda \right) ^{2\alpha _\xi }{_0F_2}\left( 1+\alpha _\xi
,3/2+\alpha _\xi ;-\frac{\alpha \xi (v\cdot k)^2}{3\pi \Lambda ^2}\right), 
\]
where $v$ is the quark velocity, $\Lambda $ is a QCD mass scale and $\alpha
_\xi \equiv \alpha (2+\xi)/3\pi $. It is totally reliable in the infrared
limit and accounts for soft-gluon corrections to the fermion in internal
loops.
\end{abstract}

\pacs{11.10Jj, 11.15Tk, 11.30Ly}

\narrowtext

\section{INTRODUCTION}


The QCD quark Lagrangian ($N$ flavors) is endowed with a higher symmetry in
the limit of equal quark velocity which applies even when the quark masses
are different\cite{IW}. Thus it generalizes the old $U(N)\times U(N)$
supersymmetry\cite{DSS} for the equal mass case. Provided that the momentum
transfer to the gluons is not much greater than the QCD scale one can
thereby deduce a number of relations between transition amplitudes, which
seem to be borne out by experiment. It has become customary\cite{G} to
attribute a velocity $v$ to the constituent heavy quark so that the momentum
of the quark in a bound state is written $p=mv+k$, where $k$ denotes the
residual quark momentum, itself associated with the light material that
makes up the hadron. As a result one can show that the ``free'' quark
propagator in the $m\rightarrow\infty$ limit is simply given by $S =
(1+\gamma\cdot v)/2v\cdot k$ and one can use this in subsequent leading 
order calculations of various matrix elements.

In this paper we would like to show that one can improve on $S$ by taking
account of soft gluon corrections. The result of the dressing is to provide
a propagator which contains the characteristic QCD scale $\Lambda$ and which
coincides with the free one in the limit of vanishing gluon coupling $\alpha
= g^2/4\pi$. Thus this propagator applies just as well to all the quarks in
the heavy mass limit\cite{BL} and does not jeopardise the prevailing higher 
symmetry. In order to derive it we use the gauge technique\cite{SDS} for QCD, 
which is known to be a reliable method in the infrared and 
ultraviolet limit\cite{DKP}. 
The technique produces a self-consistent equation for the quark spectral
function in any gauge, from which the propagator follows\cite{DW}. In the
next section we set out the velocity projector decomposition of the
propagator. Next we derive the effective vertex for soft gluons and finally
we solve the equation in question, obtaining the result quoted in the
abstract; there we also compare the result with QED where the scale $\Lambda$
is missing.

\section{Velocity projections}


When one substitutes $p=mv+k$ in the free quark propagator, the resulting
expression, 
\begin{equation}
S(p) = \frac{1}{m(\gamma\cdot v -1)+\gamma\cdot k},
\end{equation}
has to be taken in the limit $m\rightarrow\infty$ in order to discern the
resulting (leading) dependence on four-velocity $v$ (with $v^2=1$.) Now any
quark matrix $M$ for a particular flavor can be decomposed into projections
using $P_\pm = (1\pm\gamma\cdot v)/2$ according to 
\[
M = P_+M_{++}P_+ + P_+M_{+-}P_- + P_-M_{-+}P_+ + P_-M_{--}P_-. 
\]
This has the effect of resolving the $4\times 4$ matrix into four separate $%
2\times 2$ matrices: 
\[
M \Rightarrow \left( 
\begin{array}{cc}
M_{++} & M_{+-} \\ 
M_{-+} & M_{--}
\end{array}
\right). 
\]
In this basis, 
\[
v \Rightarrow \left( 
\begin{array}{cc}
1 & 0 \\ 
0 & -1
\end{array}
\right), \qquad \gamma_\mu \Rightarrow \left( 
\begin{array}{cc}
v_\mu & \gamma_\mu \\ 
\gamma_\mu & -v_\mu
\end{array}
\right). 
\]
In particular the inverse free propagator, resolves to 
\begin{equation}
S^{-1}(p) = m(\gamma\cdot v-1) + \gamma\cdot k \Rightarrow \left( 
\begin{array}{cc}
k\cdot v & \gamma\cdot k \\ 
\gamma\cdot k & -k\cdot v-2m
\end{array}
\right),
\end{equation}
and correspondingly, 
\begin{equation}
S(p) \Rightarrow \frac{1}{k^2 + (k\cdot v)^2 +2mk\cdot v} \left( 
\begin{array}{cc}
k\cdot v+2m & \gamma\cdot k \\ 
\gamma\cdot k & -k\cdot v
\end{array}
\right).
\end{equation}
Thus, up to order $1/m^2$, the free $S$ decomposes into 
\begin{equation}
S(p) \Rightarrow \frac{1}{k\cdot v}\left( 
\begin{array}{cc}
1-\frac{k^2}{2mk\cdot v} & \frac{\gamma\cdot k}{2m} \\ 
\frac{\gamma\cdot k}{2m} & -\frac{k\cdot v}{2m}
\end{array}
\right)+O(\frac{1}{m^2}),
\end{equation}
from which one infers that the leading large component is $S_{++}\sim
1/k\cdot v$.

Let us now consider the full propagator in a covariant gauge, which is best
written in the Lehmann-Kallen spectral form for our purposes: 
\begin{equation}
S(p) = \int \frac{\rho(W)\,dW}{\gamma\cdot p-W+i\eta\epsilon(W)}, \qquad
\int dW \equiv (\int_{-\infty}^{-m} + \int_m^\infty) dW,
\end{equation}
for an ordinary sort of particle. In the heavy quark limit we anticipate
that the negative energy cut is `far away' and that the main contribution
from soft gluons will arise in the vicinity of $W=m$, to within a region of
order $\Lambda$. The free propagator is of course obtained just by setting $%
\rho(W) = \delta(W-m)$ above. Put $W = m+\omega$, $p=mv+k$ and take velocity
projections as in (4), to obtain 
\begin{equation}
S(p)\Rightarrow\int\frac{\rho(\omega)\,d\omega}{2m(v\cdot
k-\omega)+k^2-\omega^2} \left( 
\begin{array}{cc}
2m+k\cdot v+\omega & \gamma\cdot k \\ 
\gamma\cdot k & -k\cdot v+\omega
\end{array}
\right).
\end{equation}
We see that the dressed propagator is still dominated by its $S_{++}$
component, which assumes a very simple form, $\int d\omega \rho(\omega)
/(v\cdot k-\omega)$ despite the inclusion of QCD interactions. More
generally, to order $1/m$, we get 
\[
S \Rightarrow \int\frac{\rho(\omega)\,d\omega}{v\cdot k-\omega} \left( 
\begin{array}{cc}
1+\frac{(v\cdot k)^2-k^2}{2m(v\cdot k-\omega)} & \frac{\gamma\cdot k}{2m} \\ 
\frac{\gamma\cdot k}{2m} & \frac{\omega-v\cdot k}{2m}
\end{array}
\right) + O(\frac{1}{m^2}). 
\]

\section{Application of the gauge technique}


The next stage involves solution of the Dyson-Schwinger (D-S) equation for
the propagator, while taking cognizance of the (longitudinal) Ward-Takahashi
identity, 
\begin{equation}
(p-p^{\prime})^\mu S(p^{\prime})\Gamma_\mu(p^{\prime},p)S(p) = S(p^{\prime})
- S(p).
\end{equation}
because it can lead to a self-consistent equation for $S$, if we ignore
certain transverse terms in the vertex $\Gamma_\mu$. If one were to
incorporate the transverse Takahashi identity\cite{TT} as well, one would in
effect be solving the full field theory; but the transverse identity---in
anything but two dimensions\cite{DT}---brings in other vertices leading to a
system of equations which is actually not closed, unless one makes further
drastic truncations\cite{K}. Alternatively if one knew the full solution of
the D-S equation in any particular gauge, one would be able to determine it
in any other gauge via the Landau-Khalatnikov-Zumino gauge covariance\cite
{LKZ} relations.

We have none of these luxuries. The gauge technique does its best to solve
the equation (7) in the form stated, while making sure that the
singularities in the non-truncated Green function are properly included. It
does not solve the inverse form of the equation (7) because that would give
a linear relation between $\Gamma$ and $S^{-1}$ and produce a difficult {\em %
nonlinear} equation for the inverse propagator; besides which, it is not
obvious how to handle the heavy quark limit for the inverse propagator---which
is dominated by its $S_{--}^{-1}$ projection, conversely to $S_{++}$.
The gauge technique starts off with the obvious solution to the vertex, 
\begin{equation}
S(p^{\prime})\Gamma_\mu^\|(p^{\prime},p)S(p)=\int dW\rho(W)\frac{1}{%
\gamma\cdot p^{\prime}-W}\gamma_\mu \frac{1}{\gamma\cdot p-W}
\end{equation}
as a weighted mass integral. One can readily check that (8) obeys (7)
automatically, but of course (8) is subject to transverse additions $%
\Gamma^\bot$; these are unknown unless one has some knowledge about them
through perturbation theory\cite{P} or examines equations for higher-order
Green functions\cite{DZ} or makes use of the transverse identity\cite{TT},
which is essentially equivalent. It is worth pointing out that such
transverse terms are soft, vanishing with the vector meson momentum. For
that reason the gauge technique is a clearly reliable tool in the infrared
limit, though it is also gauge-covariant in the ultraviolet regime\cite{DKP}
as it happens; it is only at intermediate energies that transverse
corrections to (8) play an important role.

Returning to heavy quarks, let us expand the solution (8) in powers of $1/m$
by writing it as 
\begin{equation}
S(p^{\prime})\Gamma_\mu^\|(p^{\prime},p)S(p)=\int d\omega\rho(\omega) \frac{1%
}{m(\gamma\cdot v-1)+\gamma\cdot k^{\prime}-\omega}\gamma_\mu \frac{1}{%
m(\gamma\cdot v-1)+\gamma\cdot k-\omega}
\end{equation}
and taking velocity projections: 
\begin{equation}
\Rightarrow \int \frac{\rho(\omega)\,d\omega}{(v\cdot
k^{\prime}-\omega)(v\cdot k-\omega)} \left( 
\begin{array}{cc}
v_\mu\left[ 1+\frac{(v\cdot k^{\prime})^2-k^{\prime 2}}{2m(v\cdot
k^{\prime}-\omega)}+\frac{(v\cdot k)^2-k^2}{2m(v\cdot k-\omega)} \right] +%
\frac{\gamma\cdot k^{\prime}\gamma_\mu +\gamma_\mu\gamma\cdot k}{2m} & \frac{%
v_\mu\gamma\cdot k + \gamma_\mu(\omega-k\cdot v)}{2m} \\ 
\frac{\gamma\cdot k'\cdot v_\mu + \gamma_\mu(\omega-k'\cdot v)}{2m} & 0
\end{array}
\right)
\end{equation}
up to order $1/m^2$. Note that representations (8) and (10) are {\em exact}
for small $p-p^{\prime}$ in the complete Green function (7), for the reasons
we have already given.

The next step is to use the approximation (8) in the D-S equation, which we
write in the renormalized form, 
\begin{equation}
Z^{-1}=(\gamma\cdot p-m+\delta m)S(p)+i\frac{g^2}{(2\pi)^4}\frac{\lambda^i}{2%
}\int d^4qS(p)\Gamma_\mu(p,p-q)S(p)\gamma_\nu D^{\mu\nu}(q)\frac{\lambda^i}{2%
} .
\end{equation}
Recalling the connection, $Z^{-1} = \int \rho(W)dW$, the spectral form of
the equation is, 
\begin{equation}
\int \frac {\rho(W)\,dW}{\gamma\cdot p-W} [W-m+\delta m + \Sigma(p,W)] = 0,
\end{equation}
where 
\begin{equation}
\Sigma(p,W) = i\frac{g^2}{(2\pi)^4}\frac{\lambda^i}{2}\int d^4q \gamma_\mu 
\frac{1}{\gamma.(p-q)-W}\gamma_\nu D^{\mu\nu}(q)\frac{\lambda^i}{2}
\end{equation}
is the self-energy for a quark of mass $W$ due to gluons in first order
perturbation theory.

At this point we carry out the heavy quark expansion and take velocity
projections to arrive at 
\begin{equation}
\int\frac{\rho(\omega)\,d\omega}{v\cdot k-\omega}\left[ (\omega+\delta m)
\left( 
\begin{array}{cc}
1+\frac{(v\cdot k)^2-k^2}{2m(v\cdot k-\omega)} & \frac{\gamma\cdot k}{2m} \\ 
\frac{\gamma\cdot k}{2m} & \frac{\omega-v\cdot k}{2m}
\end{array}
\right) + \left( 
\begin{array}{cc}
\Sigma_{++}(v\cdot k,\omega) & \Sigma_{+-}(v\cdot k,\omega) \\ 
\Sigma_{-+}(v\cdot k,\omega) & \Sigma_{--}(v\cdot k,\omega)
\end{array}
\right)\right] = 0,
\end{equation}
where, after summing over colours (hence the factor of 4/3),, 
\begin{equation}
\Sigma(v\cdot k,\omega)\Rightarrow i\frac{4g^2/3}{(2\pi)^4}\int \frac{%
d^4q\,D^{\mu\nu}(q)}{v\cdot (k-q)-\omega} \left( 
\begin{array}{cc}
v_\mu v_\nu \left[ 1+O(1/m)\right] & v_\mu\gamma_\nu +O(1/m) \\ 
\frac{\gamma\cdot kv_\mu v_\nu}{2m}+\frac{(\omega-k\cdot v)\gamma_\mu v_\nu}{%
2m} & \frac{\gamma\cdot k v_\mu \gamma_\nu}{2m}-\frac{(\omega-k\cdot
v)\gamma_\mu \gamma_\nu}{2m}
\end{array}
\right)
\end{equation}
and up to order $(1/m)$.

\section{The spectral equation}


To make any further progress and determine the spectral function $\rho$ and
thence the propagator, we need to make some further
approximations/assumptions about the behaviour of the gluon. It is generally
accepted that the gluons are massless so that the propagator $D(q)$ is at
least as singular as $1/q^2$; it is also known that in the ultraviolet
regime this is subject to well-defined logarthmic damping; the behaviour for
small $q^2$, where the strong force enslaves colour, is more mysterious and
there have been suggestions that $D(q)$ could be as singular as $1/q^4$,
that it plateaus or even that one should not be using QCD at all but an
effective field theory incorporating chiral symmetry with real mesons. What
is certain is the occurrence of a mass scale $\Lambda$ demarcating the
ultraviolet from the infrared regime of $D$. As we are only interested in
soft gluon effects on the heavy quark lines, we will adopt a gluon
propagator which implies masslessness, which cuts off in the ultraviolet and
which introduces the fundamental QCD mass scale. For our purposes it is
enough to use an effective 
\begin{equation}
D_{\mu\nu}(q) = (-\eta_{\mu\nu} +\xi \frac{q_\mu q_\nu}{q^2}) \frac{\Lambda^2%
}{q^2(\Lambda^2-q^2)},
\end{equation}
knowing its limitations full well. It incorporates the main things we want
and also includes a covariant gauge parameter $\xi$. If other readers wish
to modify $D$ with a more sophisticated and perhaps more realistic
expression, they can repeat our calculations below; while that is sure to
alter the precise form of our answers, we believe it will not affect the
main features of our results in a very significant way.

Returning to the largest component of (15), we have to consider the spectral
equation, 
\begin{equation}
0 = \int \frac{\rho(\omega)\,d\omega}{v\cdot k-\omega} \left[ \omega +\delta
m + \Sigma_{++}(v\cdot k,\omega) \right],
\end{equation}
\begin{equation}
\Sigma_{++}(v\cdot k,\omega) = i\frac{4g^2/3}{(2\pi)^4} \int d^4q \frac{\xi
(v.q)^2/q^2 - 1}{v.(k-q)-\omega} \left( \frac{1}{q^2} - \frac{1}{%
q^2-\Lambda^2}\right).
\end{equation}
A straightforward but messy calculation gives 
\begin{eqnarray}
\Sigma_{++}(\omega,\omega^{\prime})&=&\frac{4\alpha}{3\pi}\left[
(\omega-\omega^{\prime})(1+\frac{\xi}{2}-\xi\frac{(\omega-\omega^{\prime})^2%
}{\Lambda^2}) \ln \frac{2(\omega-\omega^{\prime})}{\Lambda}+\frac{1}{4}%
\xi(\omega-\omega^{\prime})  + \right.  \nonumber \\
& & \left. \frac{1}{2}(1-\xi\frac{(\omega-\omega^{\prime})^2}{\Lambda^2}) 
\sqrt{(\omega-\omega^{\prime})^2 - \Lambda^2} \ln \frac {\omega-\omega^{%
\prime}-\sqrt{(\omega-\omega^{\prime})^2 - \Lambda^2}} {\omega-\omega^{%
\prime}+\sqrt{(\omega-\omega^{\prime})^2 - \Lambda^2}}\right],
\end{eqnarray}
to leading order and in a general gauge. First we solve the spectral
equation (17) 
\begin{equation}
0=\int d\omega^{\prime}\,\rho(\omega^{\prime})\left[ \frac{%
\omega^{\prime}+\delta m}{\omega-\omega^{\prime}} +\frac{4\alpha}{3\pi}%
\left( \ln\frac{2(\omega-\omega^{\prime})}{\Lambda} + \frac{1}{2}\sqrt{%
(\omega-\omega^{\prime})^2 - \Lambda^2} \ln \frac {\omega-\omega^{\prime}-%
\sqrt{(\omega-\omega^{\prime})^2- \Lambda^2}} {\omega-\omega^{\prime}+\sqrt{%
(\omega-\omega^{\prime})^2 - \Lambda^2}}\right)\right]
\end{equation}
in the Fermi-Feynman gauge $\xi=0$ to discover what is going on. By taking
the imaginary part of (20), we obtain 
\begin{equation}
0 = (\omega+\delta m - 2\alpha\Lambda/3)\rho(\omega) +
(4\alpha/3\pi)\int_\omega^\infty \rho(\omega^{\prime})\,d\omega^{\prime}.
\end{equation}
This has the solution 
\[
\rho(\omega) \propto (\omega+\delta m-2\alpha\Lambda/3)^{-1-4\alpha/3\pi} 
\]
and, since the self-mass is given by 
\[
\int(\omega +\delta m)\rho(\omega)\,d\omega=0, 
\]
this fixes $\delta m = 2\alpha\Lambda/3$. It makes good sense, being governed 
by the QCD mass scale and gluon coupling. One last matter is the
proportionality factor: we must ensure that $\rho(\omega)$ reduces to $%
\delta(\omega)$ when $\alpha\rightarrow 0$. Hence we choose the overall
constant so that the result for the spectral function for $\xi=0$ is neat
and compact, namely 
\begin{equation}
\rho_{\xi = 0}(\omega) = \frac{1}{\omega\Gamma(-4\alpha/3\pi)} \left( \frac{%
\Lambda}{\omega} \right) ^{-4\alpha/3\pi}.
\end{equation}
A bonus of this choice is that the heavy quark propagator simplifies to the
elegant non-perturbative expression, 
\begin{equation}
S_{\xi=0}(v\cdot k) = \Gamma(1+\frac{4\alpha}{3\pi})\frac{1+\gamma\cdot v}{%
2v\cdot k} \left( -\frac{\Lambda}{v\cdot k} \right) ^{-4\alpha/3\pi}.
\end{equation}
In the limit $\alpha \rightarrow 0$ one recovers the free result $1/(v\cdot
k)$ for $S$.

Now we turn to the general gauge $\xi$. Noting that the $\xi$-dependent part
of $\Sigma_{++}(v\cdot k,\omega)$ vanishes at the threshold $v\cdot k=\omega$%
, the gauge-dependence of $\rho$ arises purely from the imaginary part of $%
\Sigma_{++}$. In this way (21) gets modified to 
\begin{equation}
0 = (\omega+\delta m -2\alpha/3\pi)\rho(\omega) + \frac{4\alpha}{3\pi}
\int_\omega^\infty d\omega^{\prime}\,\rho(\omega^{\prime}) \left[ 1+\frac{\xi%
}{2}-\xi\frac{(\omega-\omega^{\prime})^2}{\Lambda^2}\right].
\end{equation}
Once again the self-mass condition requires $\delta m=2\alpha\Lambda/3$,
which is satisfyingly {\em gauge-independent}. The resulting integral 
equation for
the spectral function is a little bit harder to solve now; nevertheless one
may establish that it reduces to a generalised hypergeometric function: 
\begin{equation}
\rho(\omega) = \frac{1}{\omega\Gamma(-\alpha_\xi)} \left( \frac{\Lambda}{%
\omega} \right) ^{-2\alpha_\xi} {_0F_2}\left( 1+\alpha_\xi, 3/2+\alpha_\xi;-%
\frac{\alpha\xi\omega^2}{3\pi\Lambda^2}\right),
\end{equation}
where $\alpha_\xi\equiv\alpha(2+\xi)/3\pi$. Thereupon\cite{fn} the heavy
quark propagator becomes 
\begin{equation}
S(v\cdot k) = \Gamma(1+2\alpha_\xi)\frac{1+\gamma\cdot v}{2v\cdot k} \left( -%
\frac{\Lambda}{v\cdot k} \right) ^{-2\alpha_\xi} {_0F_2}\left( 1+\alpha_\xi,
3/2+\alpha_\xi;-\frac{\alpha\xi(v\cdot k)^2}{3\pi\Lambda^2}\right) + O(1/m).
\end{equation}
This is universal to all the quarks and could be used to estimate the gluon
corrections in loops which result from dressing  fermion lines and their
vertices. However, a word of caution: the result (26) does not take account
of gluon self-interactions; those will somehow need to be included
separately in heavy quark calculations.

There is one further test of our work. One needs to verify that to order $%
1/m $ the other, `small component' sectors in the propagator velocity
projections are correctly determined by (25), because they are fixed in
terms of the leading $\rho $. We have indeed checked this out: the $S_{+-}$
sector produces precisely the same equation as (24), while the $S_{--}$
sector is nothing but the self-mass condition, $\int (\omega +\delta m)\rho
(\omega )\,d\omega =0$, which we have already settled\cite{fn2}.

Finally it is worth comparing the answers with the gauge technique solutions
for scalar QED say. Those solutions do not have the benefit of an intrinsic
cut-off; rather the source mass itself acts as the cutoff and the results
read, 
\[
m^2\rho (W)=\frac{(W^2/m^2-1)^{-1-2a_\xi}}{\Gamma (-2a_\xi )}{%
_2F_1(-a_\xi ,1-a_\xi ;-2a_\xi ;1-\frac{W^2}{m^2});\qquad a_\xi =(\xi
+2)\alpha /4\pi }
\]
\[
m^2S(p)=\Gamma (1+a_\xi )\Gamma (2+a_\xi ){_2F_1}(1+a_\xi ,2+a_\xi
;2;p^2/m^2)
\]
We notice a strong similarity with (25) and (26), which becomes greater when
one substitutes $p=mv$ and expands to order $1/m$. The only change is that $m$
takes the place of $\Lambda $ as the argument of the hypergeometric function.
It only remains to obtain the non-leading behaviour of $S$ in the various
sectors. This entails solving the spectral equation up to order $1/m$ and is
where our use of the approximations (9) and (16) start to look a bit suspect
because they are connected with gluons which carry off appreciable momentum.
It is a nice subject for future research since it portrays the
mass-dependence of the heavy quark Lagrangian.

\acknowledgments
We thank the University Research Committee for providing a small grant
during 1996 which enabled this collaboration to take place.


\begin{references}
\bibitem{fn} We have made use of the Hilbert transform, $\int_0^\infty 
\frac{\omega'^\gamma\, d\omega^{\prime}}{\omega-\omega^{\prime}} =
\pi\cot(\pi\gamma)\,  \omega^\gamma$ and noted that $_0F_2$ has $\omega^2$
 as its argument, in deriving the result quoted.
\bibitem{fn2}  The renormalization factor $Z^{-1}$ comes out
infinite in our case.
\end{references}
\end{document}